\documentclass[emulateapj,natbib]{emulateapj}

\begin{document}

\title{Resolving the  HD 100546 Protoplanetary System with the \textrm{Gemini Planet Imager}: Evidence for Multiple Forming, Accreting Planets}
\author{
Thayne Currie\altaffilmark{1}, 
Ryan Cloutier\altaffilmark{2,8,9},
Sean Brittain\altaffilmark{3},
Carol Grady\altaffilmark{4,10},
Adam Burrows\altaffilmark{5},
Takayuki Muto\altaffilmark{6},
Scott J. Kenyon\altaffilmark{7},
Marc J. Kuchner\altaffilmark{4}
}
\altaffiltext{1}{National Astronomical Observatory of Japan, Subaru Telescope}
\altaffiltext{2}{Department of Astronomy and Astrophysics, University of Toronto}
\altaffiltext{3}{Department of Physics \& Astronomy, Clemson University}
\altaffiltext{4}{Exoplanets and Stellar Astrophysics Laboratory, NASA-Goddard Space Flight Center}
\altaffiltext{5}{Department of Astrophysics Sciences, Princeton University}
\altaffiltext{6}{Division of Liberal Arts, Kogakuin University}
\altaffiltext{7}{Smithsonian Astrophysical Observatory}
\altaffiltext{8}{Centre for Planetary Sciences, University of Toronto}
\altaffiltext{9}{Institut de recherche sur les exoplan\`{e}tes, Universit\'{e} de Montr\'{e}al}
\altaffiltext{10}{Eureka Scientific}
\begin{abstract}
We report \textit{Gemini Planet Imager} H band high-contrast imaging/integral field spectroscopy and polarimetry of the HD 100546, a 10 $Myr$-old early-type star recently confirmed to host a thermal infrared bright (super)jovian protoplanet at wide separation, HD 100546 b.  We resolve the inner disk cavity in polarized light, recover the thermal-infrared (IR) bright arm, and identify one additional spiral arm.  We easily recover HD 100546 b and show that much of its emission plausibly originates from an unresolved, point source.   The point source component of HD 100546 b has extremely red infrared colors compared to field brown dwarfs, qualitatively similar to young cloudy superjovian planets; however, these colors may instead indicate that HD 100546 b is still accreting material from a circumplanetary disk.     Additionally, we identify a second point source-like peak at $r_{proj}$ $\sim$ 14 AU, located just interior to or at inner disk wall consistent with being a 10--20 $M_{J}$ candidate second protoplanet-- ``HD 100546 c" --  and lying within a weakly polarized region of the disk but along an extension of the thermal IR bright spiral arm.    Alternatively, it is equally plausible that this feature is a weakly polarized but locally bright region of the inner disk wall.  Astrometric monitoring of this feature over the next 2 years and emission line measurements could confirm its status as a protoplanet, rotating disk hot spot that is possibly a signpost of a protoplanet, or a stationary emission source from within the disk.  
\end{abstract}
\keywords{planetary systems, stars: early-type, stars: individual: HD 100546} 
\section{Introduction}
The $\sim$ 10 $Myr$-old early-type star HD 100546 provides an excellent laboratory for studying newly-formed/forming jovian planets and the interaction between  protoplanetary disks and embedded (proto)planets.  Its disk is cleared of small dust out to $\approx$ 10--15 AU and contains numerous (planet induced?) spiral arms \citep{Grady2005,Boccaletti2013,Currie2014c}.   Independent studies confirm the existence of a $\approx$ 1--10 $M_{J}$ protoplanet, HD 100546 b, at $\approx$ r $\sim$ 0\farcs{}5 (50 AU) \citep{Currie2014c,Quanz2013,Quanz2015}.  Additionally, rovibrational CO gas monitoring suggests the presence of an unseen infant planet at/near the inner disk cavity wall \citep{Brittain2014}.   HD 100546's circumstellar environment is likely an evolutionary precursor to systems of multiple superjovian mass planets \citep[e.g. HR 8799,][]{Marois2008,Marois2010}.

Near-infrared (Near-IR; 1-2.5$\mu m$) photometry/spectroscopy could constrain HD 100546 b's emission source and colors.  But efforts with conventional adaptive optics (AO)  imaging have yielded only non-detections \citep{Boccaletti2013,Quanz2015}.   However, \textit{extreme} (Strehl Ratio $\gtrsim$ 80--90\%) AO imaging such as with the \textit{Gemini Planet Imager} \citep[GPI;][]{Macintosh2014} yields contrasts 10--100 times deeper at HD 100546 b-like angular separations.    Near-IR extreme AO data can also improve our knowledge of HD 100546's disk via polarimetry and image/place strict limits on the predicted second companion.  

In this Letter, we present  GPI $H$-band coronagraphic integral field spectroscopy (IFS) and polarimetry of HD 100546\footnote{This paper's analysis is limited to \textit{photometry} extracted for any (candidate) planets.  A separate, later work will analyze spectra (Cloutier et al., in prep.)}.  We resolve the inner disk cavity, reimage the thermal IR-bright spiral arm, and recover the HD 100546 b protoplanet.   We identify  a bright emission source at 13 AU along the thermal IR bright spiral arm,  consistent with a second protoplanet (a candidate ``HD 100546 c") or a weakly polarized and particularly luminous region of the inner disk wall.  
\\
\section{Observations}
\subsection{$H$ band IFS Data}
We observed HD 100546 with GPI  on 18 January 2015 under exceptional (0\farcs{}3 in the optical) seeing conditions in $H$ band ($\lambda_{o}$ = 1.6 $\mu$m, $R$ = 44--49; 0\farcs{}014166 pixel$^{-1}$) behind the apodized pupil Lyot coronagraph 
($r_{mask}$ $\sim$ 0\farcs{}12) and in angular differential imaging mode \citep[ADI;][]{Marois2006}.   Due to a slight off-centering of the mask edge, our inner working angle on the south side of the field is smaller ($\sim$ 0\farcs{}11) than the north side ($\sim$ 0\farcs{}13).  Our observations consist of 55 co-added 59.5 second frames, yielding a total integration time of $\approx$ 55 minutes and a field rotation of  24$^\circ$, or 1.3 (5) $\lambda$/D at 0\farcs{}13 (0\farcs{}5).   
Basic data reduction utilized the \textit{Gemini Data Reduction Pipeline, version 1.2.1} \citep{Perrin2014}.

For point-spread function (PSF) subtraction, we used the \textit{adaptive, locally optimized combination of images} algorithm \citep[A-LOCI][]{Currie2012,Currie2014b} considering a moving pixel mask and 
exploring a range of algorithm settings:  the rotation gap criterion ($\delta$), 
the optimization area ($N_\mathrm{A}$), the \textit{singular value decomposition} (SVD) cutoff (\textit{SVD$_\mathrm{lim}$} ).   Our previous detection of HD 100546 b favored algorithm settings including a large rotation gap, high SVD cutoff and pixel masking, since HD 100546 b is embedded in strong, azimuthally-varying disk signal \citep{Currie2014c}.  A small rotation gap allows us to access regions down to the coronagraph mask edge.   As a compromise, we incorporated a moving pixel mask and a small rotation gap ($\delta$ = 0.5-0.6) but an aggressive SVD cutoff (\textit{SVD$_\mathrm{lim}$} = 0.001--0.01), the latter similar to including the first few principle components in KLIP-based PSF subtraction \citep[e.g.][]{Soummer2012}.    We constructed a final data cube from a median-combination of PSF-subtracted cubes and collapse the cube to yield a band-averaged image.   

To separately assess the HD 100546 disk structure in total intensity, we rotated each data cube to north-up, median-combining the cubes using a $\sigma$-clipped mean, collapsing the cube to form a final image, and spatially filtered this image using a 15x15 moving box.

\subsection{$H$ band Coronagraphic Polarimetry (PI)}
We downloaded public HD 100546 $H$ band coronagraphic polarmetry obtained during GPI commissioning on 12 December 2013, consisting of four 58.2 second exposures at four waveplate position angles with the same coronagraph mask as used with the IFS data, although better centered.  
For basic processing, we subtracted a dark frame, determined the position of the lenslet spots,  destriped the image, and interpolated over bad pixels.   We constructed a registered polarization data cube consisting of two orthogonal polarizations, subtracting the mean stellar and instrumental polarization as estimated from the signal within the mask.   We use double differencing to clean the polarization pairs, rotate each image to north up and combine the sequence to produce a total intensity image and the $Q$, $U$, and $V$ polarized intensity images.   

Combining the $Q$ and $U$ images as in \citet{Avenhaus2014} yields linear polarized intensity images along the perpendicular and parallel directions (\textit{P$_{\perp}$} and \textit{P$_{//}$}) .  We equate \textit{P$_{\perp}$} with the true polarized intensity.  To conservatively estimate the polarization measurement uncertainty, we equate \textit{P$_{//}$} with noise.


\section{Detections}
\subsection{PI Data and Spatially-Filtered IFS Data}
\begin{figure}
\centering
\includegraphics[scale=0.4]{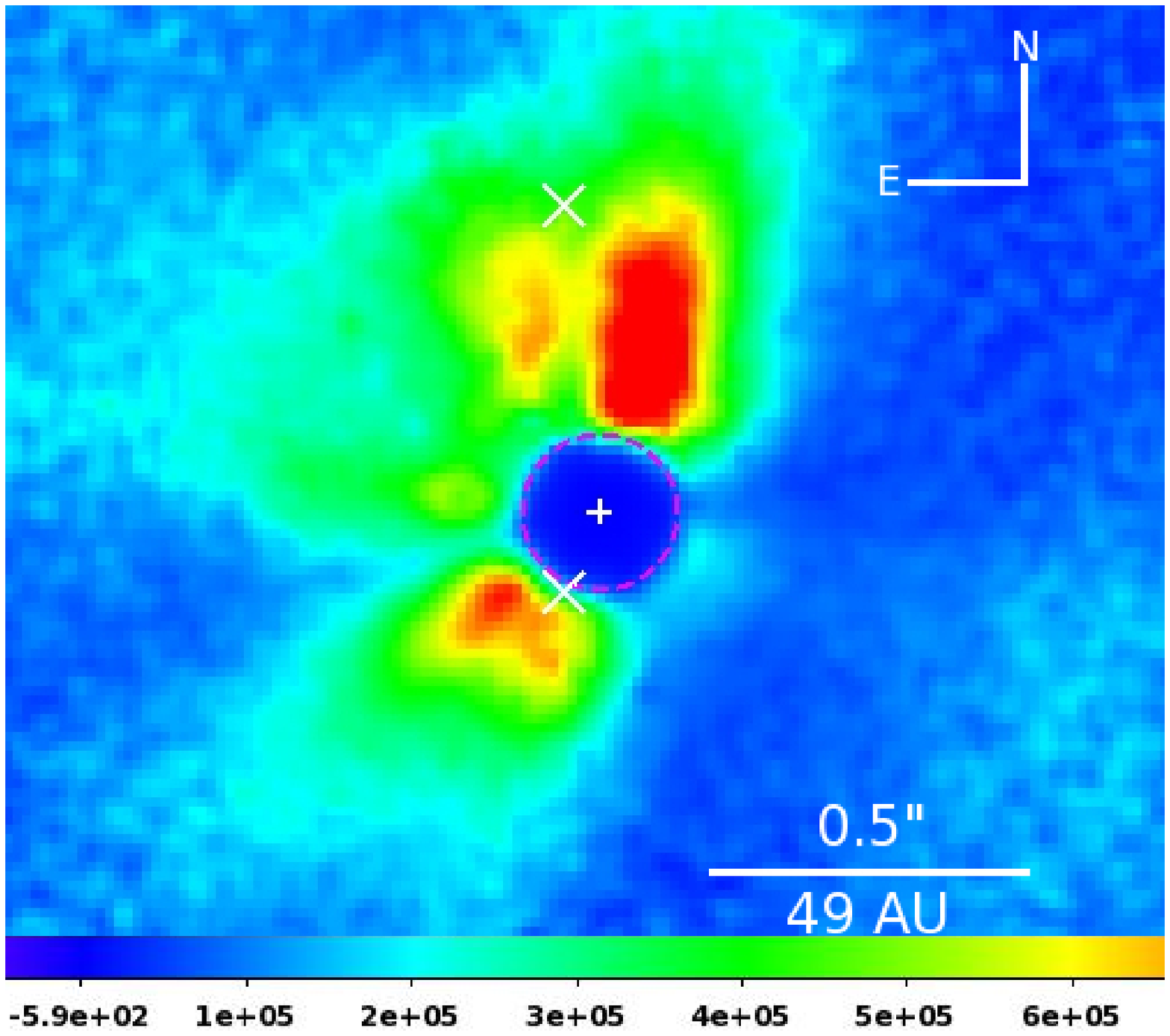}
\includegraphics[scale=0.4]{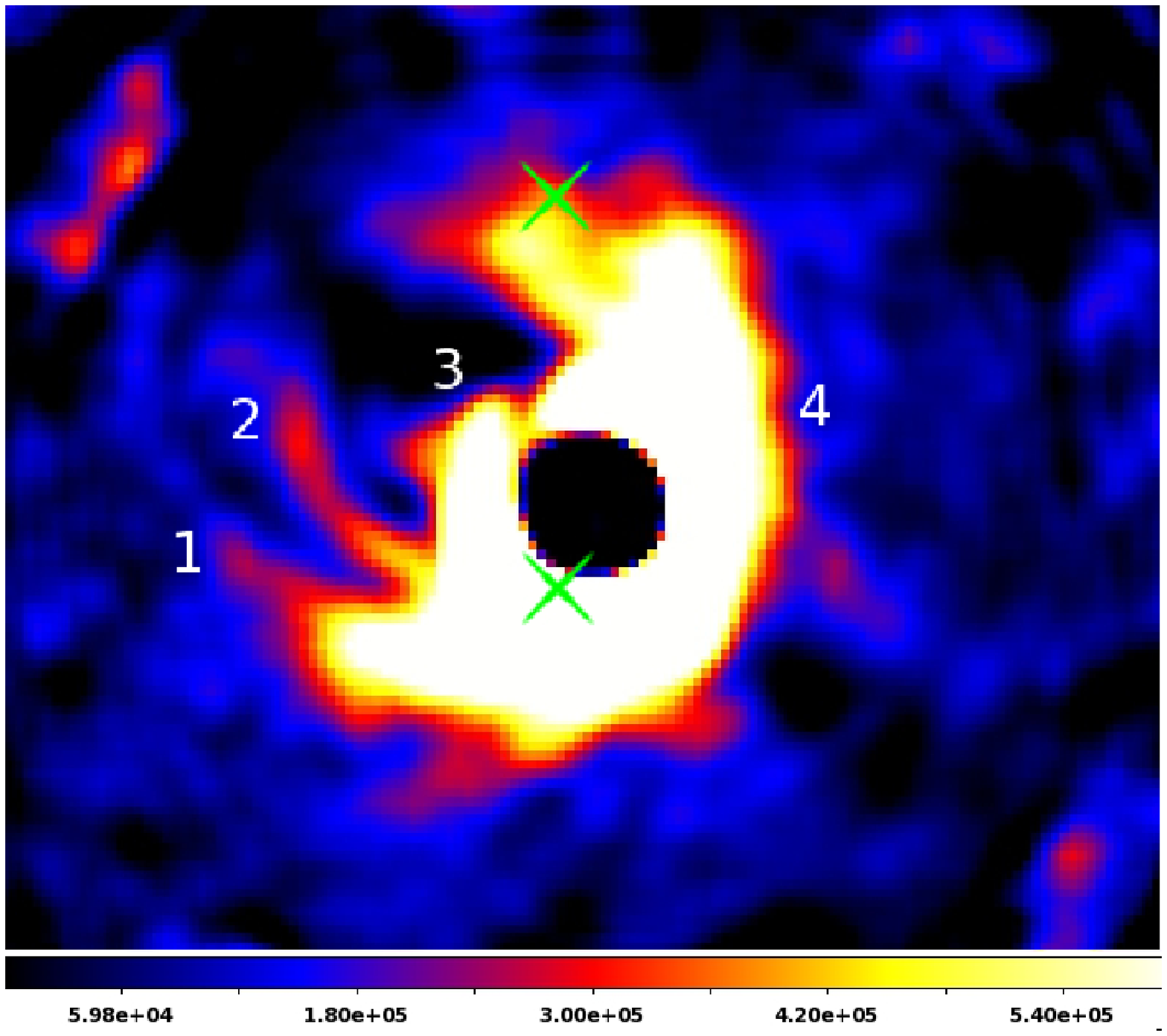}
\caption{HD 100546 stellocentric distance-scaled disk images;  an \textit{x} denotes the positions of  HD 100546 b and the candidate ``HD 100546 c".  (left) Polarized intensity image, showing the inner working angle (magenta dashed line) and the star's position (cross).   The polarized intensity drops interior to 0\farcs{}15-0\farcs{}2.  The dark blue regions the west of the visible disk identify the dark lane \citep{Grady2001,Grady2005}.  (right) Spatially-filtered, wavelength collapsed IFS image, showing the thermal IR bright spiral-like feature (1), a second spiral (2), the position of the proposed spiral identified by \citet{Avenhaus2014} (3), and the western rim of the visible disk (4).}
\label{diskimage}
\end{figure}

Figure \ref{diskimage} shows the reduced PI image (left), the image rescaled by its stellocentric distance squared (center), and the spatially-filtered wavelength-collapsed IFS image (right).    The disk's  PI maxima is exterior to our inner working angle ($r$ $\sim$ 0\farcs{}12) at $r$ $\sim$ 0\farcs{}15, 0\farcs{}16 and PA $\sim$ 340$^{\circ}$ and 50$^{\circ}$ for the northwest and southeast sides, respectively  \citep[see also][]{Avenhaus2014}.    In contrast to previous results, the northwest region closest to the star has the highest polarization, roughly 1.5x that of the southeast region.   

The spatially-filtered IFS image recovers the thermal IR bright arm seen \citep{Currie2014c} plus a second (fainter) candidate spiral arm located clockwise from the thermal IR-bright one.   \citet{Avenhaus2014} identify another clockwise wrapping spiral at small angular separations: our data recover this feature, although its nature (a spiral or a spiral-like feature made by a lack of dust emission at PA $\sim$ 45$^{\circ}$) is unclear.   In both polarized and total intensity, the disk signal drops precipitously beyond $\sim$ 0\farcs{}5--0\farcs{}55.

\subsection{PSF Subtracted IFS Data}
\begin{figure}
\centering
\includegraphics[scale=0.4,trim=37mm 0mm 37mm 0mm,clip]{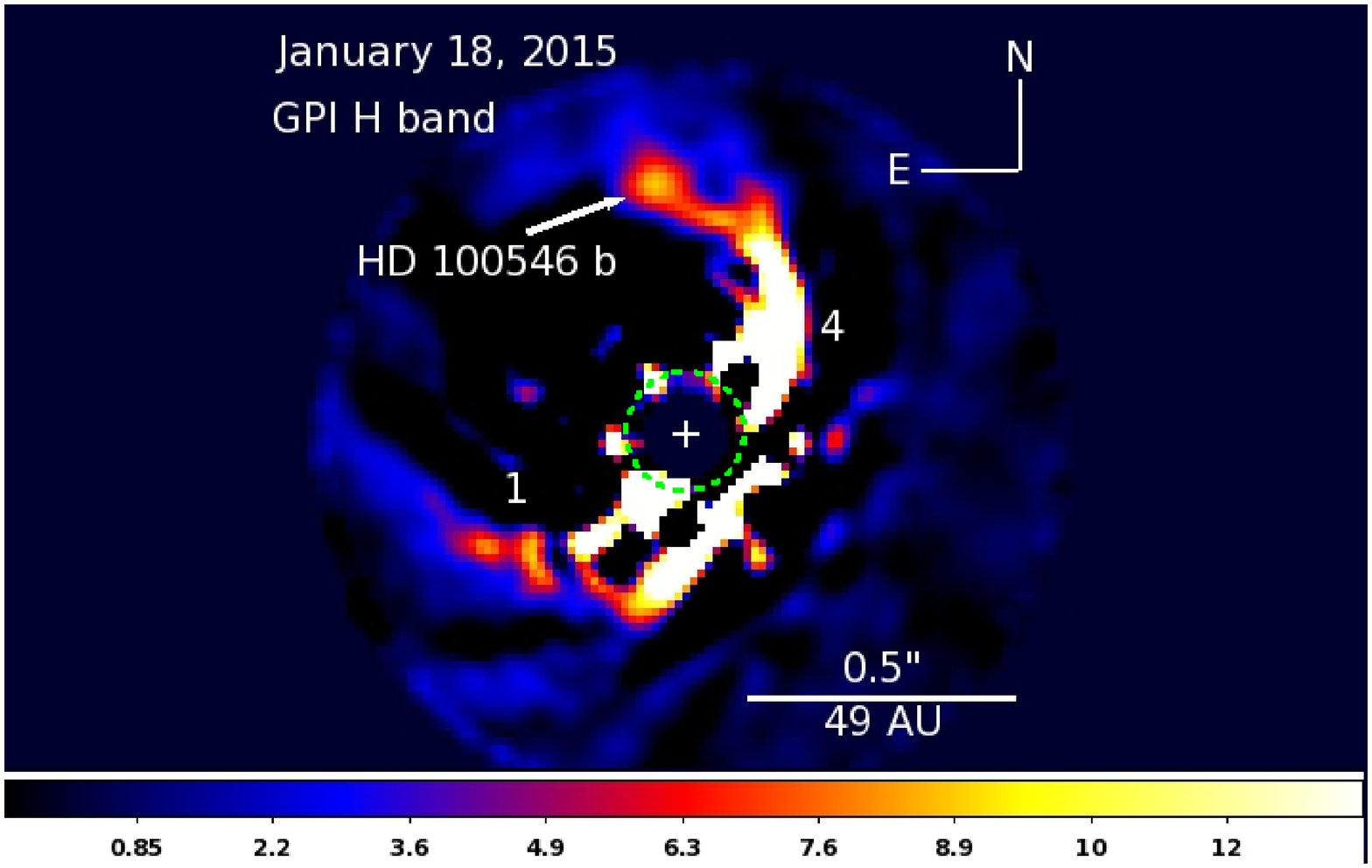}
\includegraphics[scale=0.4,trim=37mm 0mm 37mm 0mm,clip]{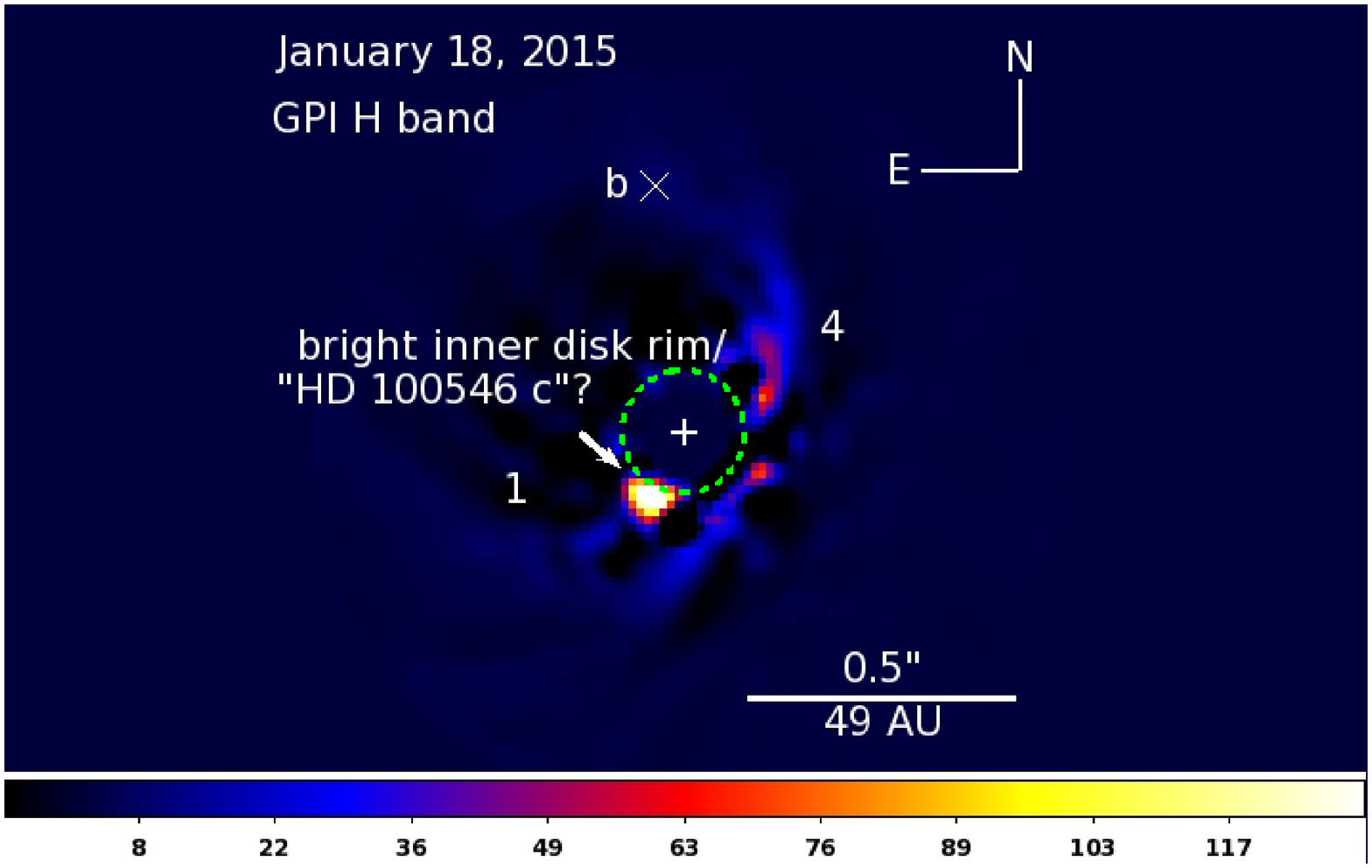}\\
\includegraphics[scale=0.4,trim=37mm 17.25mm 37mm 0mm,clip]{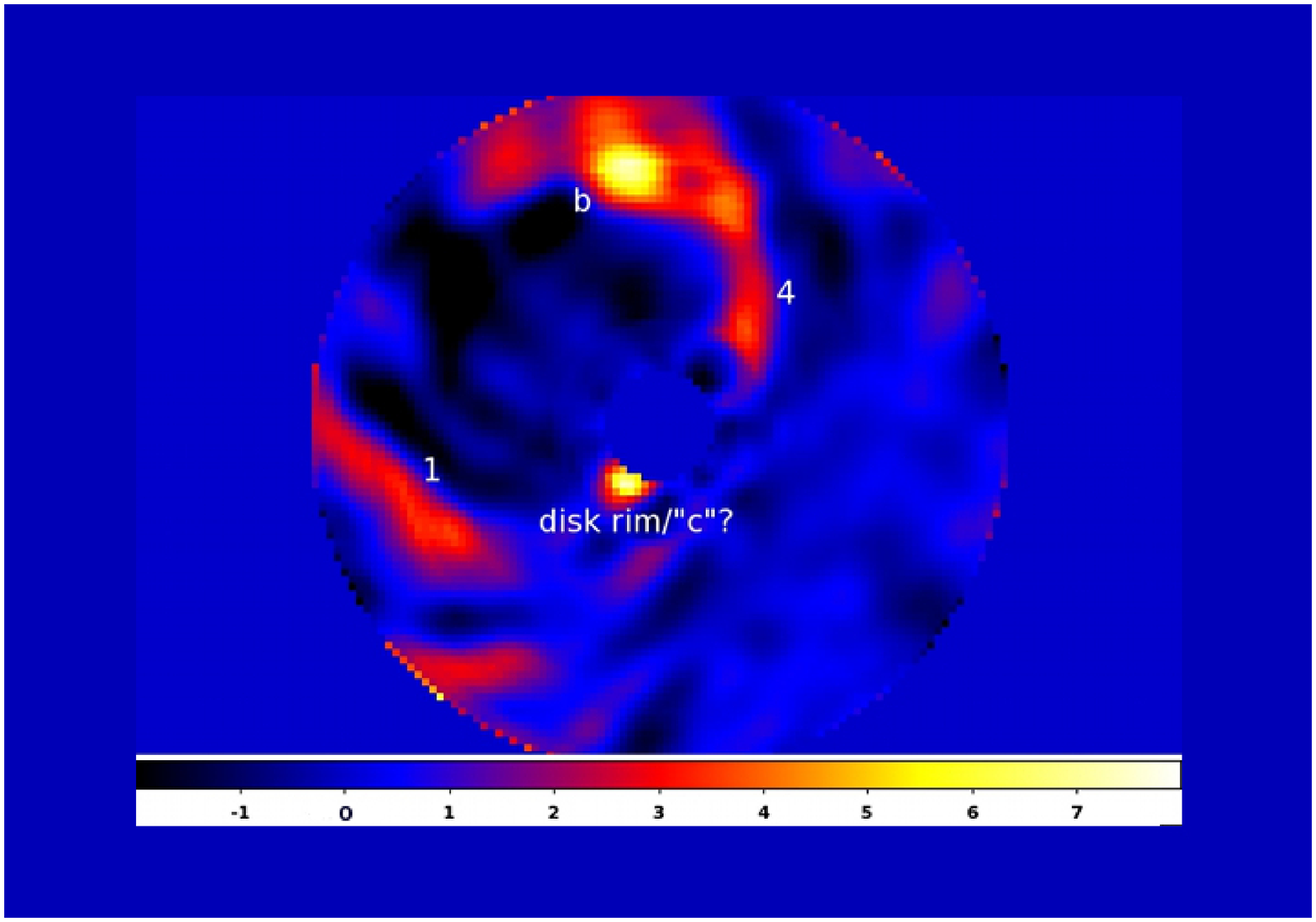}
\includegraphics[scale=0.4,trim=37mm 0mm 37mm 0mm,clip]{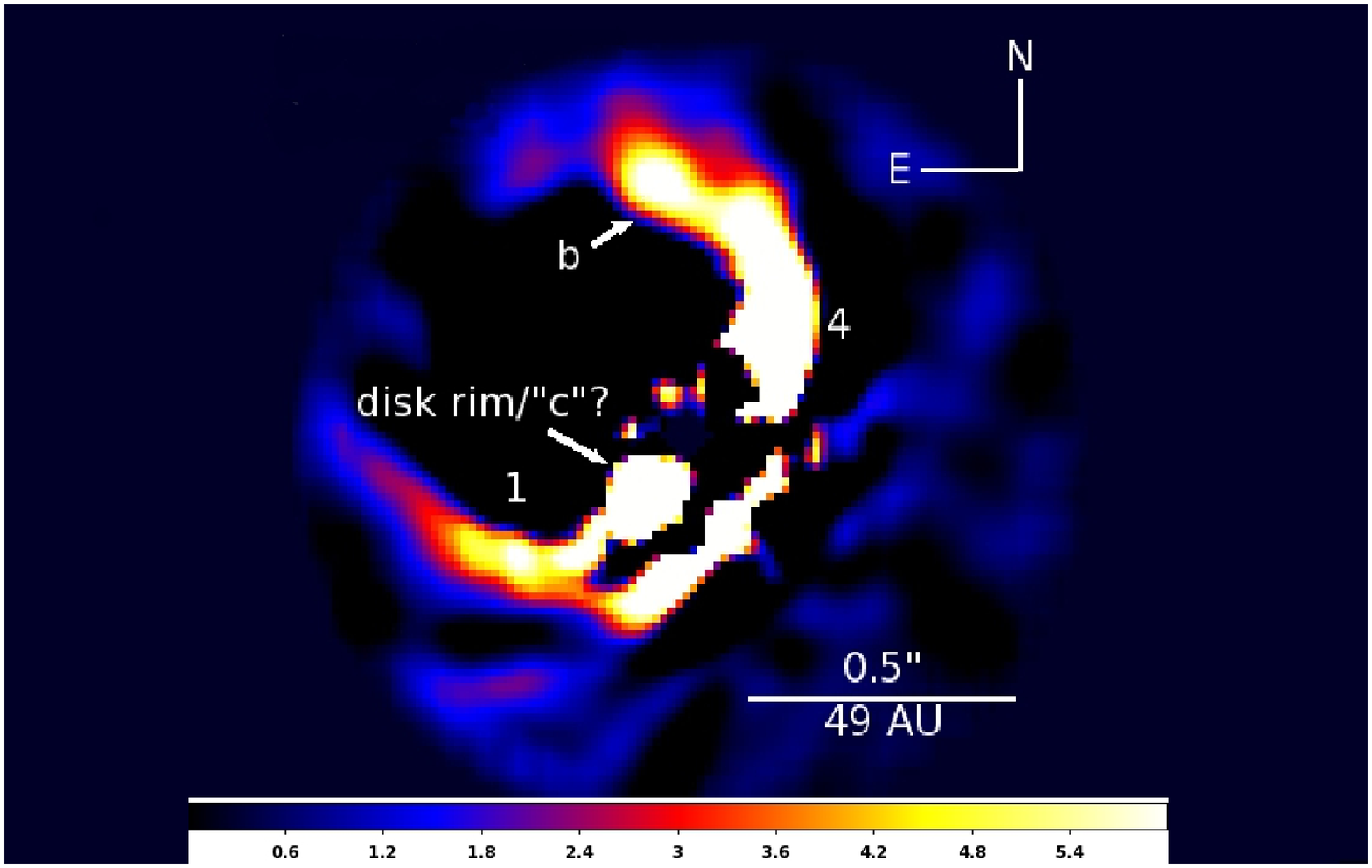}
\caption{(Top)  PSF-subtracted, wavelength-collapsed image of HD 100546 from IFS data, showing HD 100546 b (left) and a second point source-like feature that may be a bright disk rim or a candidate planet ``HD 100546 c" (right; image stretch multiplied by ten).  HD 100546 b and "c" are about 5.5\% and 9.4\% as bright as the rms of the speckle halo+disk in the raw data, about a factor of 2--3 brighter relative to the halo noise than HR 8799 bcde in $H$-band extreme-AO data from Project 1640 \citep{Oppenheimer2013}.  (Bottom) (left) SNR map for the IFS image and (right) the positions of the (candidate) protoplanets relative to disk material in a box-car smoothed (5x5 pixels) image.   
  }
\label{planetimage}
\end{figure}

The A-LOCI-reduced IFS data easily recover HD 100546 b at a location ($r$ $\sim$ 0\farcs{}469 $\pm$ 0\farcs{}012 PA = 7.04$^{\circ}$ $\pm$ 1.39$^{\circ}$ ) consistent with previous detections (Figure \ref{planetimage}).  Adopting the conservative signal-to-noise ratio calculations (SNR) described in \citet{Currie2014c} \footnote{All SNR estimates are corrected for final sample sizes \citep[][]{Mawet2014}.}, the protoplanet appears in the wavelength-collapsed image at SNR $\sim$ 7.  However, its true significance compared to residual speckle noise is far higher, since the +/- 2-3 $\sigma$ wings of the probability density distribution of the convolved image at HD 100546 b's location are dominated by real astrophysical features (e.g. the thermal IR-bright spiral and its self-subtraction footprints), not residual speckles.  Defining the noise from a ring radius at least 3 pixels more distant,  a better assessment of residual speckle noise (just beyond the disk edge in polarized and scattered light), HD 100546 b has a SNR $>$ 11.  

The IFS data reveals evidence for additional protoplanets.  
We recover the thermal IR-bright spiral arm albeit partially subtracted such that only the spine is visible.   
Additionally, our image and SNR map identify a second peak  at $r$ $\sim$ 0\farcs{}131 $\pm$ 0\farcs{}009, PA $\sim$ 150.85$^{\circ}$ $\pm$ 1.98$^{\circ}$ with a SNR of $\sim$ 7, consistent with being a second protoplanet, ``HD 100546 c", lying along an extension of the thermal IR-bright arm.   Multiple additional reductions spanning a range of algorithm space confirm our interpretation that this peak is a real astrophysical feature.   
\section{Analysis}
\subsection{The HD 100546 Disk in Polarized Light}
\begin{figure}
\centering
\includegraphics[scale=0.55]{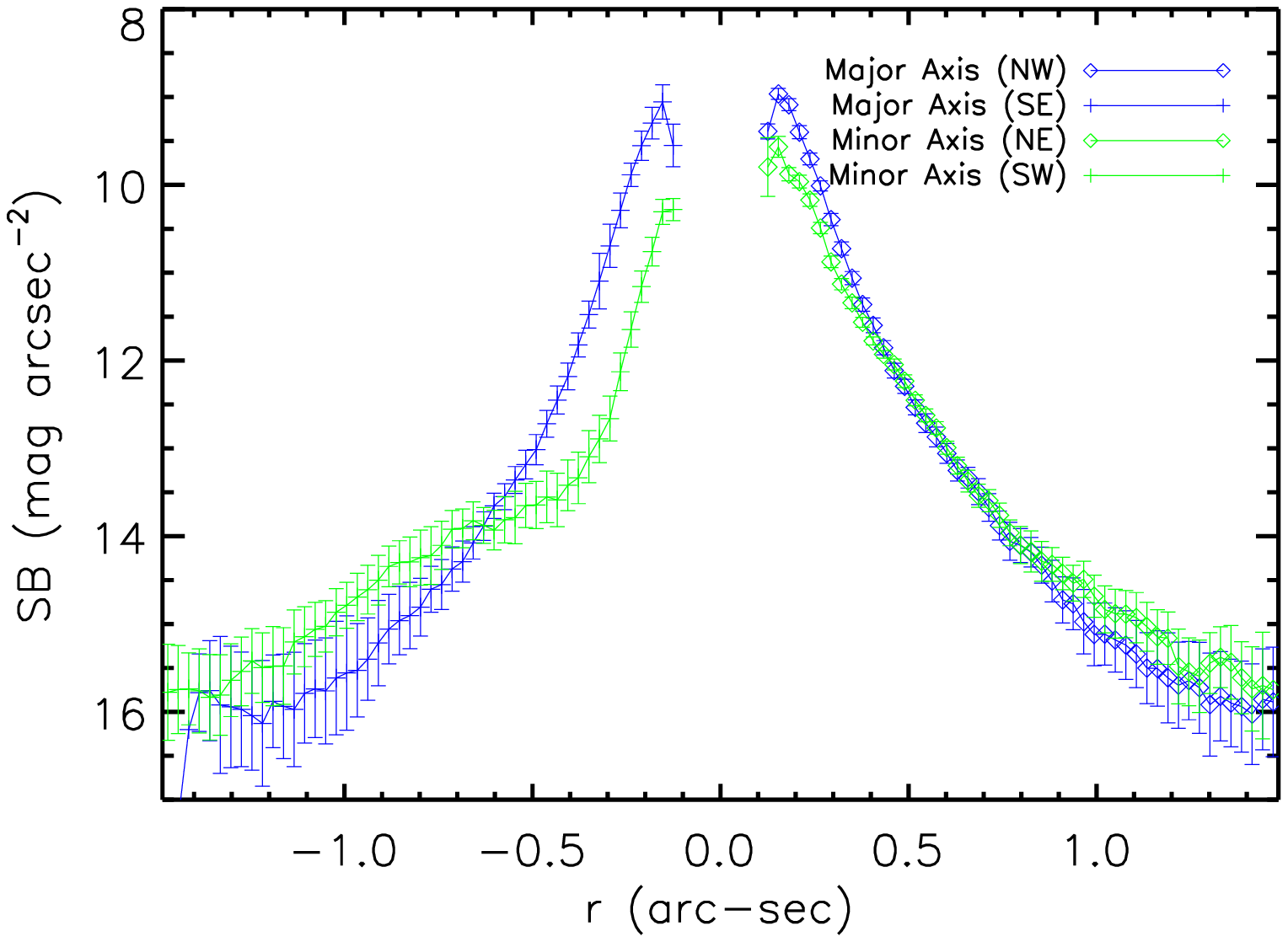}
\includegraphics[scale=0.3]{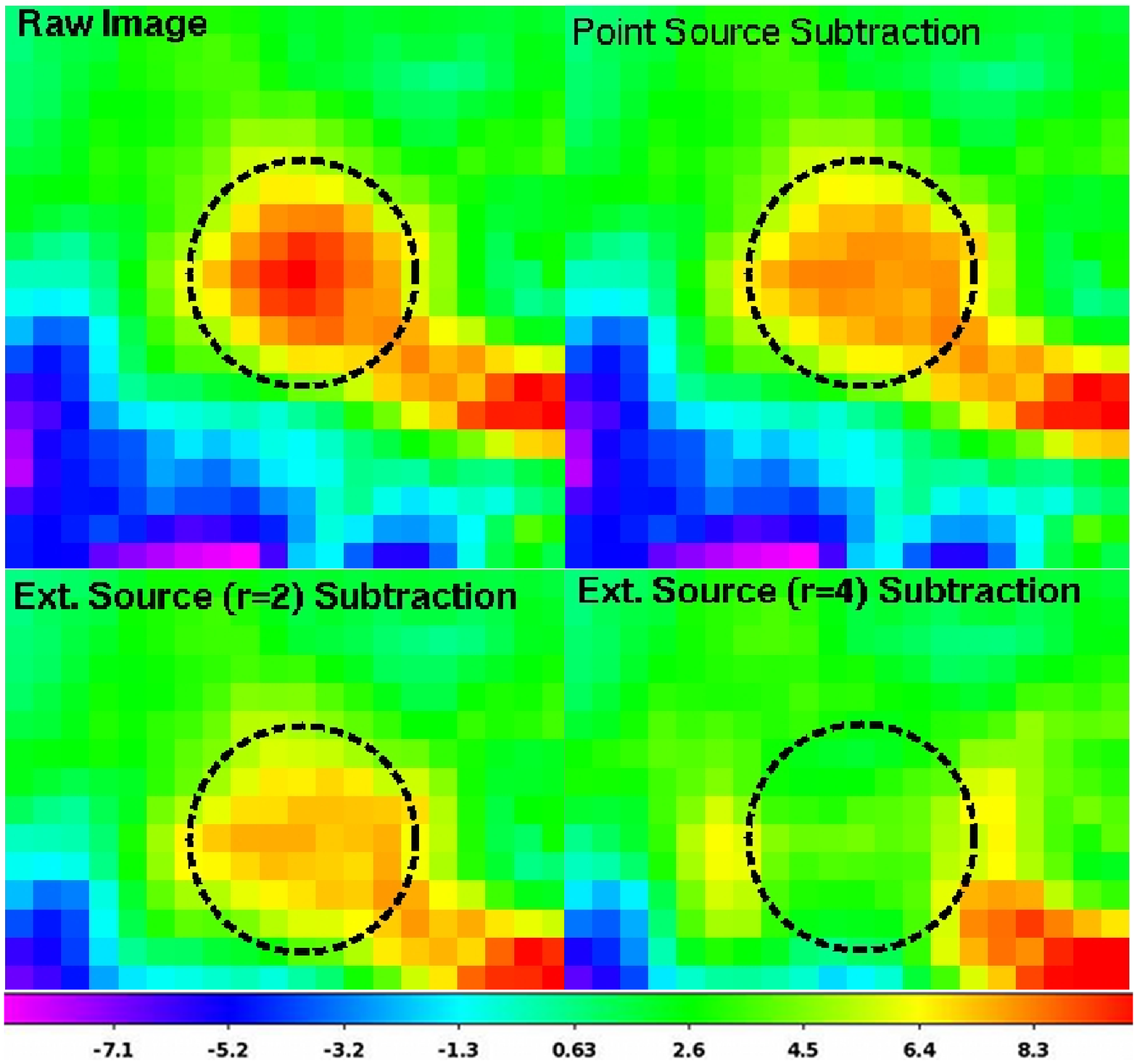}
\caption{(left) PI surface brightness profile of the HD 100546 disk computed in 10 degree wedges as in \citet{Avenhaus2014}.  (right) Close-up of final IFS image showing region around HD 100546 b and the subtraction of HD 100546 b assuming point source and 2 or 4-pixel wide extended sources.   The background disk emission covers a circular region (dotted circle).}
\label{disksb}
\end{figure}

Our polarimetry data clarify the morphology of the HD 100546 disk.    Following \citet{Avenhaus2014} and \citet{Quanz2011}, we determined the disk's surface brightness (SB) profile along the major and minor axes (Figure \ref{disksb}, left panel), fixing the northwest major axis brightness at 0\farcs{}15 to 9.0 mag arcsec$^{-2}$ as measured in these papers.     While our PI profiles generally agree with these studies,  polarization is \textit{lower} at $r$ $>$ 1\arcsec{}.
 In agreement with \citet{Avenhaus2014}, we see evidence for a turnover in SB at $r$ $\sim$ 0\farcs{}15--0\farcs{}16 consistent with the presence of the disk's inner cavity at a projected separation $r_{proj}$ $\approx$ 14 AU.  Following methods outlined in \citet{Avenhaus2014}, our derived disk position angle is $\approx$ 148$^{\circ}$ as measured from the location of the brightest pixels on the northwest and southeast sides or $\approx$ 154$^{\circ}$ from the ``centroid" of the brightest regions of these sides.   We cannot disentangle the stellar PSF from the disk in total intensity nor can we use the interpolation method of \citet{Perrin2014} to extract the total intensity disk signal because the disk is not confined to a ring but essentially covers 2$\pi$ radians from r $\sim$ 0\farcs{}14 to r  $\sim$ 0\farcs{}6.  Thus, we cannot derive a map of the disk fractional polarization.

There is no polarized intensity peak at the location of HD 100546 b (Figure 1a, white x); the southeast polarization peak is not coincident with but counterclockwise of the candidate companion's position.   The thermal IR bright spiral arm and other (candidate) spiral arms do not appear in polarized light.
\subsection{Characterization of HD 100546 b}
\subsubsection{HD 100546 b With a Point Source Component}
\citet{Currie2014c} find that HD 100546 b plus the surrounding circumstellar disk material can be matched by a simple gaussian whose FWHM is consistent with a slightly extended emission source; \citet{Quanz2015} model HD 100546 b's emission as originating from a point source and spatially resolved emission from the circumstellar disk, reporting photometry for the point source component.   With a smaller PSF core, GPI data allow us to better determine whether HD 100546 b is better interpreted as a point source or extended source. 

 To study HD 100546 b's morphology, we simulate the detection and annealing of signals with a variety of spatial scales using A-LOCI forward-modeling  \citep{Currie2015a}.  We derive the GPI PSF in each wavelength slice from the average of satellites in the an unsharp-masked, median-combined stack of data cubes.  We consider an unresolved point source and extended sources up to 4 pixels in radius, comparable to sizes found by \citet{Currie2014c} to be consistent with HD 100546 b's  $L^\prime$ detection and determine the proper scaling between the model PSF and the data by minimizing the determinant of the Hessian matrix at HD 100546 b's position \citep[see][]{Quanz2015}.  

Subtracting a best-fit point source at HD 100546 b yields a flat disk signal without over subtracting the disk (Figure \ref{disksb}, right panel), providing evidence that HD 100546 b likely has a point-source component.  A two-pixel radius extended source cannot easily be ruled out, but produces higher residuals along the PSF y-axis and some evidence of over subtraction (i.e. a slight depression in signal at HD 100546 b's location fainter signal).  A four-pixel radius extended source yields very strong residuals: brighter, nonphysical peaks lying just exterior the circular region defining the background disk feature.

\subsubsection{HD 100546 b's Infrared Colors and Circumplanetary Material}
Assuming HD 100546 b is a point source at its core, we can estimate the point source component's photometry and colors.  Using the best-fit point source scaling, HD 100546 b's apparent magnitude: $m_{H}$ = 19.40 $\pm$ 0.32, where the errors consider the intrinsic SNR, the uncertainty in the satellite spot calibration, and the uncertainty in the throughput correction.   Using the models of \citet[][]{Mulders2011,Mulders2013} adopted in \citet{Quanz2015}, the disk extinction at HD 100546 b's location is  $A_{H}$ = 3.4.  Assuming a distance of 97 $\pm$ 4 $pc$ \citep{Berriman1994}, the unreddened absolute magnitude is then $M_{H}$ = 11.06 $\pm$ 0.33.

\begin{figure}
\centering
\includegraphics[scale=0.1]{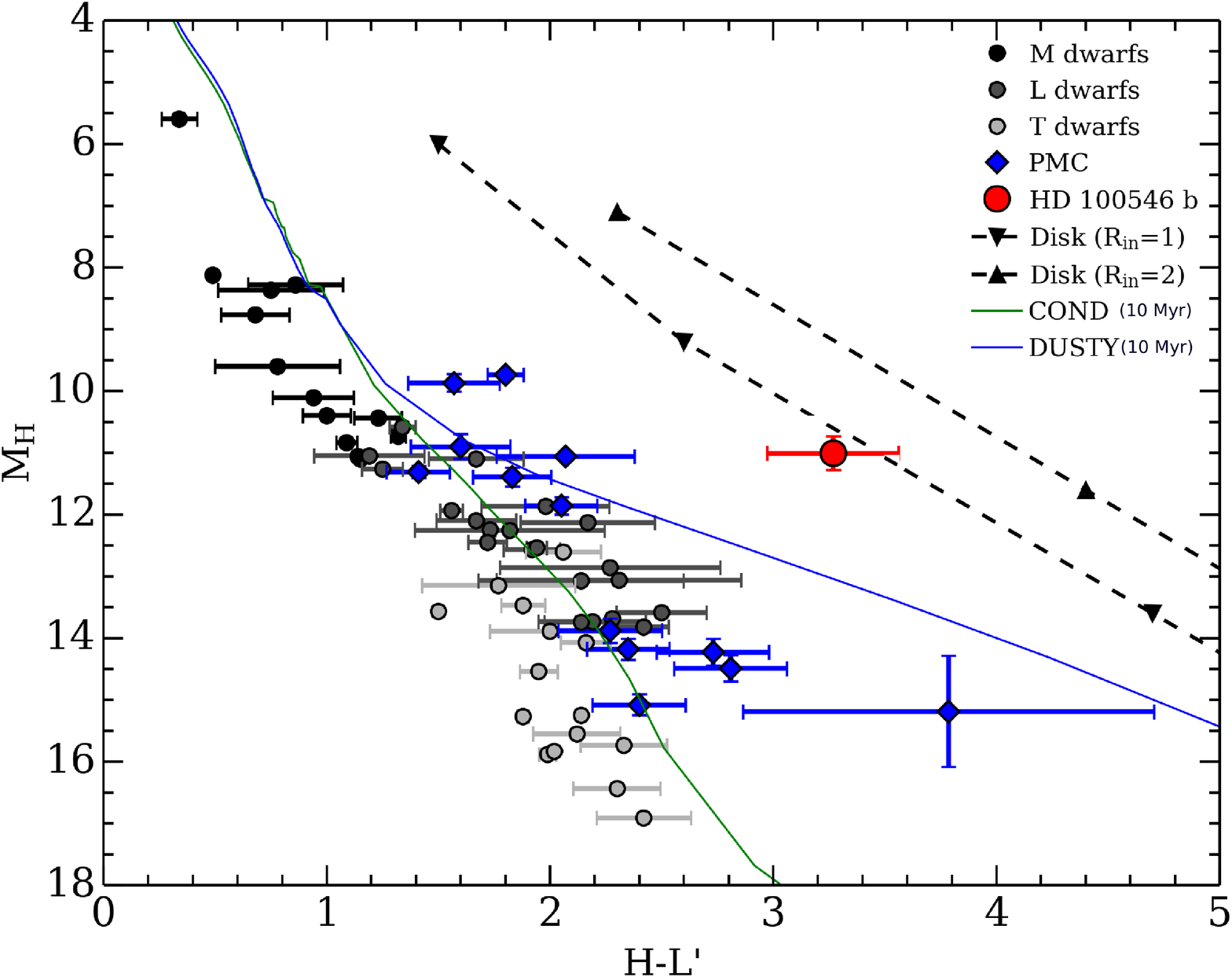}
\includegraphics[scale=0.18]{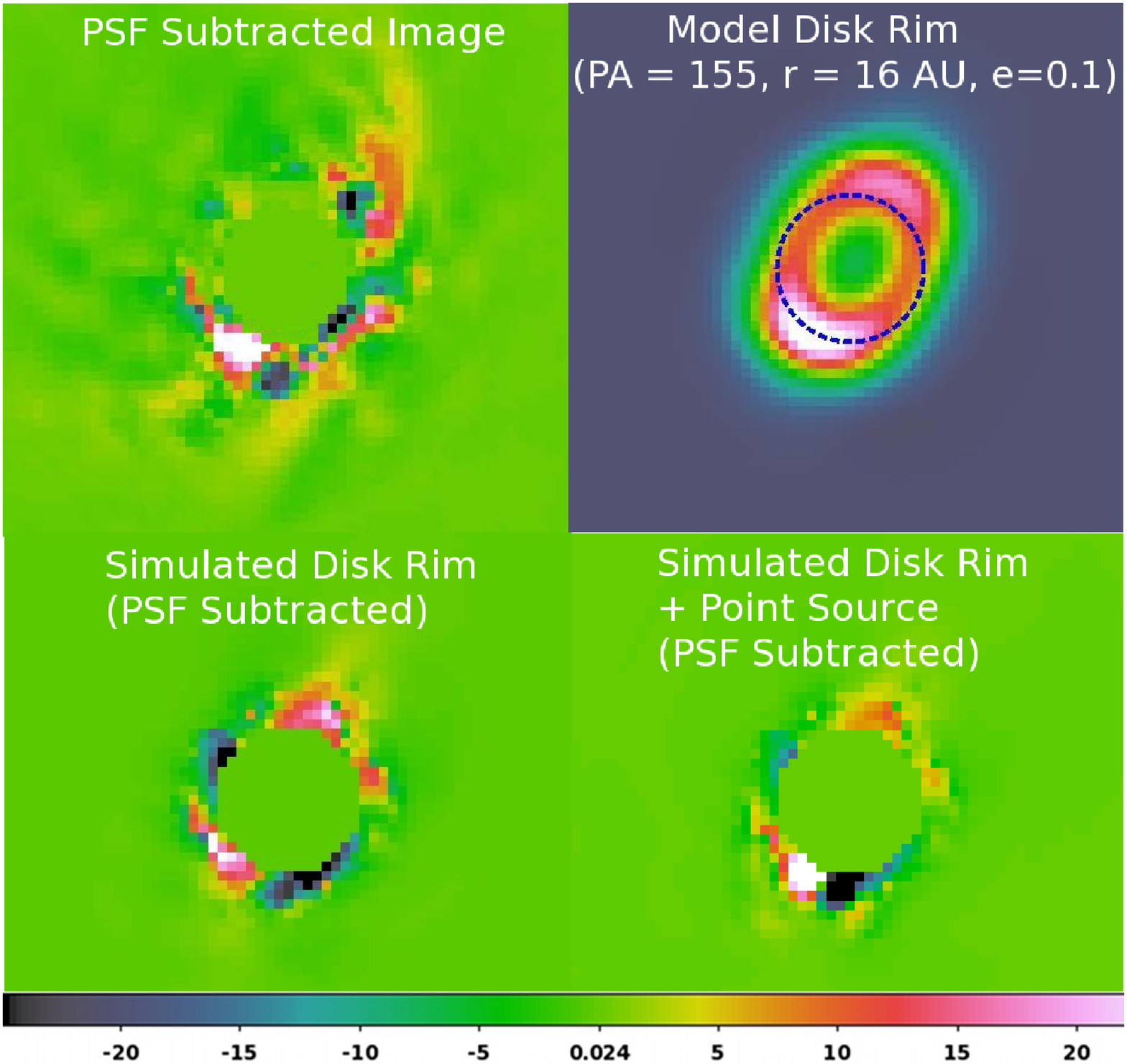}
\caption{(Left) $H$-$H$-$L^\prime$ color-magnitude diagram comparing HD 100546 b to the MLT dwarf sequence from \citet{Leggett2010}, imaged planet-mass companions (PMC), the COND and DUSTY limiting-case planet/brown dwarf atmosphere model loci \citep{Allard2001,Baraffe2003} and the loci of colors for accretion-dominated protoplanets  \citep{Zhu2015}.    PMC draw from the compilation in \citet{Currie2013} updated to include new ROXs 42Bb photometry  \citep{Currie2014a,Currie2014d}.  (Right)  PSF subtracted image and model disk rim compared to a simulated, PSF-subtracted disk rim and a rim+point source.}
\label{bcmd}
\end{figure}

   As shown in Figure \ref{bcmd}, the protoplanet's $H$ absolute magnitude is comparable to that for objects near the M/L transition, but with $H$-$L^\prime$ $\approx$ 3.2 it is redder than all but the reddest L/T transition dusty/cloudy young superjovian planet, HD 95086 b \citep{Rameau2013,Galicher2014}.   Red near-to-mid IR colors have been cited as evidence for thicker clouds than present in field brown dwarfs with similar temperatures \citep{Currie2011a}.  However, HD 100546 b lies redward of the AMES-DUSTY model locus, the rough limit of a realistic cloudy atmosphere filled with submicron-sized dust.

The infrared colors of HD 100546 b's point source component suggest the presence of a circumplanetary disk as hypothesized by \citet{Currie2014c}.   
The object is redder than GSC 06214B, a substellar companion whose accreting circumsecondary disk produces a broadband mid-IR excess \citep{Bowler2011,Bailey2013}.    The predicted diagram positions for accreting protoplanets from \citet{Zhu2015} matches HD 100546 b's position for an inner radius of 1 $R_{J}$ and an accretion rate of 3 $\times$10$^{-6}$ $M$/$M_{J}$$\times$($M_{J}$ yr$^{-1}$).   For an inner radius of 2 $R_{J}$, comparable to the estimated radii for the youngest imaged planets \citep{Currie2013,Currie2014b}, the models are about 0.75 mags too red.  
The inconsistency can be reconciled if the $H$-band emission originates from a higher temperature planetary photosphere (e.g. $T_{eff}$ $\approx$ 1000--2000 K), while the disk dominates the $L^\prime$ emission.  HD 100546 b's $H$-band photometry then implies a mass of 3--4 $M_{J}$ if the companion is newly born (e.g. 1 $Myr$ old) \citep[c.f.][]{Spiegel2012,Baraffe2003}. 

\subsection{`HD 100546 c": A Weakly Polarized Disk Feature or A Second Protoplanet?}

The putative ``HD 100546 c" requires careful scrutiny despite its high detection significance.  On one hand, its position is at an identical separation to and within 2-$\sigma$ of the position angle predicted for a second protoplanet (``HD 100546 c") orbiting HD 100546 near the inner disk wall ($r$ $\approx$ 0\farcs{}14, 130$^{\circ}$ $\pm$ 10$^{\circ}$ for 18 January 2015).   It is not the same feature as the local peak in polarized intensity; it appears \textit{interior to} the PI-inferred disk wall.
However, the candidate is located very close to the mask edge ($\Delta$r$_{``c"-mask}$ $\approx$ 0\farcs{}020):  disk emission continuing at smaller separations would be preferentially extincted on one side.  If ``HD 100546 c" is instead the brightest region of an inner disk rim which extends to r $<$ 0\farcs{}12, the mask extinction in combination with PSF annealing could make it appear like a point source.

To assess these effects, we simulated ``HD 100546 c" with forward-modeling as a partially subtracted 1) disk ring coplanar with the main disk, 2)  extended source 4 pixels wide, 3)  point source, and 4) combinations of 1--3.  For scenario 1), we constructed a grid of models using GRATER \citep{Augereau1999}, varying the position angle (145--155$^{\circ}$), eccentricity (0-0.1), density power law ($\alpha_{in}$ = 10,5) and radius (14--16 AU).  

Figure \ref{bcmd} (right panel) compares representative simulated images to the data and to the disk rim model.  The simulated rim model struggles to reproduce the shape of the candidate: the emission confined far too narrowly in angular separation and shifted clockwise.  The 4-pixel wide extended source model (not shown) fails in a similar manner.   While the simulated point source (not shown) shows far better agreement at the location of the candidate, combining its signal with that of the simulated disk rim model (especially for PA = 150-155$^\circ$) fares even better, reproducing both the main signal \textit{and} the residual trail of residual emission wrapping clockwise from the seven o'clock position to nine o'clock position.    

 Photometry for ``HD 100546 c" is broadly consistent with that expected from a young protoplanet.  If it lies interior to the disk, our forward-modeling suggests an upper limit on the point source component's apparent magnitude of 13.08 $\pm$ 0.64, comparable to a 10--20 $M_{J}$ newly-formed planet \citep{Spiegel2012,Baraffe2003}, consistent with estimates based on modeling the inner disk's morphology \citep{Mulders2013}.

\section{Discussion and Future Work}
HD 100546 b likely includes a point source component and derives some of its emission from circumplanetary material.   Its $H$-band spectrum could provide the first assessment of how a protoplanet spectrum compares to that for fully-formed planets and brown dwarfs (Cloutier et al., in prep.).   Given extreme computational challenges in forward-modeling the complex HD 100546 disk structure and our lack of knowledge about what an infant protoplanet's spectrum (likely including planet and circumplanetary signal) should look like, extracting a spectrum for the disk to compare with that of HD 100546 b and decisively interpreting the protoplanet is not feasible right now.  However, future observations utilizing a reference PSF may obviate the need for forward modeling of the disk and allow us to at least directly compare the disk's spectrum to that of the protoplanet.  

Although circumstantial evidence suggests that the putative ``HD 100546 c" may be the proposed second planet responsible for clearing out the inner disk cavity, our analysis is not conclusive since the source is close to the disk wall and the GPI coronagraph mask edge.   Alternate scenarios include a stationary, weakly polarized disk feature and an orbiting (unresolved?) disk hot spot that may or may not be induced by a planet.   If the candidate is a planet and is rapidly accreting gas, it may show up as unusually bright in $H_{\alpha}$ or Pa$\beta$. 

Orbital monitoring may also clarify the nature of this candidate.
A planet in Keplerian orbit at 13 AU around a 2.4 $M_{\odot}$ star and coplanar with the disk moves $\approx$ 12$^{\circ}$ year$^{-1}$ in the \citeauthor{Brittain2014} model.  Projected on the sky this motion is $\approx$ 2 GPI pixels over the next two years and thus may be too small to confirm orbital motion to distinguish this candidate from a stationary disk feature.  However, the predicted companion from \citeauthor{Brittain2014} should have been at an orbital phase of $\approx$ 154$^{\circ}$ on 18 January 2015.    Once its orbital phase reaches 180$^{\circ}$ it will be positioned directly \textit{behind} the disk wall in projection.  Thus, if the candidate ``HD 100546 c" is the predicted protoplanet inferred from CO rovibrational measurements, not a stationary disk feature or a planet-induced hot spot on the disk wall, it will likely disappear from view sometime in 2017.  CO monitoring and GPI multi-epoch imagery combined with wider parameter space of simulated disk rim models should soon determine the nature of this candidate
 and clarify the inner circumstellar environment of HD 100546.

\textbf{Acknowledgements} -- We thank Gijs Mulders, Wladimir Lyra, Nienke van der Marel, and the anonymous referee for helpful comments.  This work is supported through NASA Origins of Solar Systems program NNG13PB64P. 


{}

\begin{thebibliography}{}
\bibitem[Allard et al.(2001)]{Allard2001}Allard, F., Hauschildt, P., Alexander, D. R., et al., 2001, \apj, 556, 357
\bibitem[Augereau et al.(1999)]{Augereau1999}Augereau, J. C., et al. 1999, A\&A, 348, 557
\bibitem[Avenhaus et al.(2014)]{Avenhaus2014}Avenhaus, H., Quanz, S., Meyer, M. R., et al., 2014, A\&A, 790, 56
\bibitem[Bailey et al.(2013)]{Bailey2013}Bailey, V., Hinz, P., Currie, T., et al., 2013, \apj, 767, 31
\bibitem[Baraffe et al.(2003)]{Baraffe2003}Baraffe, I., Chabrier, G., Barman, T. S., et al., 2003, A\&A, 402, 701
\bibitem[Berriman et al.(1994)]{Berriman1994}Berriman, G. B., Boggess, N. W., Hauser, M. G., et al., 1994, \apj, 431, L63
\bibitem[Boccaletti et al.(2013)]{Boccaletti2013}Boccaletti, A., Pantin, E., Lagrange, A.-M., et al., 2013, A\&A, 560, L20
\bibitem[Bowler et al.(2011)]{Bowler2011}Bowler, B., Liu, M. C., Kraus, A., et al., 2011, \apj, 743, 148
\bibitem[Brittain et al.(2014)]{Brittain2014}Brittain, S., Carr, J., Najita, J., 2014, \apj, 791, 136
\bibitem[Burrows et al.(2006)]{Burrows2006}Burrows, A., Sudarsky, D., Hubeny, I., 2006, \apj, 640, 1063
\bibitem[Currie et al.(2011)]{Currie2011a}Currie, T., Burrows, A., et al., 2011, \apj, 729, 128
\bibitem[Currie et al.(2012)]{Currie2012}Currie, T., Debes, J., Rodigas, T., et al., 2012, \apj, 760, L32
\bibitem[Currie et al.(2013)]{Currie2013}Currie, T., Burrows, A., Madhusudhan, N., et al., 2013, \apj, 776, 15
\bibitem[Currie et al.(2014a)]{Currie2014a}Currie, T., Daemgen, Debes, J., et al., 2014a, \apj, 780, L30
\bibitem[Currie et al.(2014b)]{Currie2014b}Currie, T., Burrows, A., Girard, J., et al., 2014b, \apj, 795, 133
\bibitem[Currie et al.(2014c)]{Currie2014c}Currie, T., Muto, T., Kudo, T., et al., 2014c, \apj, 796, L30
\bibitem[Currie et al.(2014d)]{Currie2014d}Currie, T., Burrows, A., Daemgen, S., 2014, \apj, 787, 104
\bibitem[Currie et al.(2015)]{Currie2015a}Currie, T., Lisse, C., Kuchner, M., et al., 2015, \apj, 807, L7
\bibitem[Galicher et al.(2014)]{Galicher2014}Galicher, R., Rameau, J., Bonnefoy, M., et al., 2014, A\&A, 565, L4
\bibitem[Grady et al.(2001)]{Grady2001}Grady, C., A., Polomski, E. F., Henning, Th., et l., 2001, \aj 122, 3396
\bibitem[Grady et al.(2005)]{Grady2005}Grady, C. A., Woodgate, B., Heap, S. R., et al., 2005, \apj, 620, 470
\bibitem[Lafreni\`ere et al.(2007)]{Lafreniere2007a}Lafreni\'ere, D., Marois, C., Duyon, R., et al., 2007, \apj, 660, 770
\bibitem[Leggett et al.(2010)]{Leggett2010}Leggett, S., et al., 2010, \apj, 710, 1627
\bibitem[Macintosh et al.(2014)]{Macintosh2014}Macintosh, B., Graham, J., Ingraham, P., et al., 2014, PNAS, 111, 35
\bibitem[Marois et al.(2006)]{Marois2006}Marois, C., Lafreni\'ere, D., Duyon, R.,  al., 2006, \apj, 641, 556
\bibitem[Marois et al.(2008)]{Marois2008}Marois, C., Macintosh, B., Barman, T., et al., 2008, Science, 322, 1348
\bibitem[Marois et al.(2010)]{Marois2010}Marois, C., Zuckerman, B., Konopacky, Q., et al., 2010, Nature, 468, 1080
\bibitem[Mawet et al. (2014)]{Mawet2014}Mawet, D., Milli, J., Wahhaj, Z., et al., 2014, \apj, 792, 97
\bibitem[Mulders et al.(2011)]{Mulders2011}Mulders, G., Waters, L. B. F. M., Dominik, C., et al., 2011, A\&A, 531, 93
\bibitem[Mulders et al.(2013)]{Mulders2013}Mulders, G., Paardekooper, S.-J., Panic, O., et al., 2013, A\&A, 557, 68
\bibitem[Oppenheimer et al.(2013)]{Oppenheimer2013}Oppenheimer, R., Baranec, C., Beichman, C., et al., 2013, \apj, 768, 24
\bibitem[Perrin et al.(2014)]{Perrin2014}Perrin, M., Maire, J., Ingraham, P., et al., 2014, SPIE, 9147, 3
\bibitem[Quanz et al.(2011)]{Quanz2011}Quanz, S., Schmid, H. M., Geissler, K., et al., 2011, \apj, 738, 26
\bibitem[Quanz et al.(2013)]{Quanz2013}Quanz, S., Meyer, M. R., Kenworthy, M., et al., 2013, \apj, 766, L1
\bibitem[Quanz et al.(2015)]{Quanz2015}Quanz, S., Amara, A., Meyer, M. R., et al., 2015, \apj, 807, 64
\bibitem[Rameau et al.(2013)]{Rameau2013}Rameau, J., Chauvin, G., Lagrange, A.-M., et al., 2013, \apj, 779, L26
\bibitem[Soummer et al.(2012)]{Soummer2012}Soummer, R., Pueyo, L., Larkin, J.,  2012, \apj, 755, L28
\bibitem[Spiegel and Burrows(2012)]{Spiegel2012}Spiegel, D., Burrows, A., 2012, \apj, 745, 174
\bibitem[Zhu(2015)]{Zhu2015}Zhu, Z., 2015, \apj, 799, 16
\end{thebibliography}
\end{document}